%% file: motionCaptureMessagePassing.tex
\documentclass[a4paper,10pt]{article}
\pdfoutput=1
\usepackage[bookmarks=false]{hyperref}
\usepackage[numbers]{natbib}
\usepackage[margin=1in]{geometry}

\usepackage{graphicx}
\usepackage[english]{babel}
\usepackage{float}
\usepackage{enumitem} 
\usepackage[cmex10]{amsmath}
\usepackage{amsfonts}
\usepackage{amssymb}
\usepackage{algorithm}     
\usepackage{algpseudocode} 
\usepackage{graphicx}
\usepackage{caption}
\usepackage[caption=false,font=footnotesize]{subfig}
\usepackage{color}
\usepackage{tikz,pgfplots}
\usepackage{mathtools}
\usepackage{array,booktabs,tabularx}
\newcolumntype{C}{>{\centering\arraybackslash}X}

\newcommand{\ie}{i.e.\ }
\newcommand{\eg}{e.g.\ }
\newcommand{\Sectionref}[1]{\hyperref[#1]{Section~\ref*{#1}}}
\newcommand{\Figureref}[1]{\hyperref[#1]{Figure~\ref*{#1}}}
\newcommand{\Tableref}[1]{\hyperref[#1]{Table~\ref*{#1}}}
\newcommand{\Stepref}[1]{\hyperref[#1]{Step~\ref*{#1}}}
\newcommand{\Algorithmref}[1]{\hyperref[#1]{Algorithm~\ref*{#1}}}

\DeclareMathOperator*\argmin{arg \, min \,}

\newcommand{\Transp}{\mathsf{T}}

\usepgfplotslibrary{external}
\pgfplotsset{compat=newest}
\pgfplotsset{plot coordinates/math parser=false}
\newlength\figureheight
\newlength\figurewidth
\usetikzlibrary{arrows}
\usetikzlibrary{calc}
\usetikzlibrary{matrix}
\usetikzlibrary{backgrounds}
\usetikzlibrary{fit}
\usetikzlibrary{positioning}
\usetikzlibrary{patterns}
\usetikzlibrary{shapes}
\usepgfplotslibrary{groupplots}
\usetikzlibrary{calc}
\usepackage{color}
\DeclareMathOperator*{\minimize}{minimize}
\DeclareMathOperator*{\maximize}{maximize}
\DeclareMathOperator*{\st}{subj. \ to}

\DeclareMathOperator*{\minimum}{min}

\DeclareMathOperator*{\parent}{par}

\DeclareMathOperator*{\children}{ch}

\DeclareMathOperator*{\leaves}{leaves}

\newcommand{\indexOneClique}{a}
\newcommand{\indexTwoClique}{b}
\newcommand{\indexThreeClique}{b}
\newcommand{\separatorSet}{A}
\newcommand{\nCliques}{N_C}
\newcommand{\cliqueTreeIndex}{c}

\newcounter{remcount}
\newtheorem{rem}[remcount]{Remark}

\begin{document}
\input{coverArXiv.tex}

\title{\textbf{A Scalable and Distributed Solution to the Inertial Motion Capture Problem}}

\author{Manon~Kok$^\star$, Sina~Khoshfetrat~Pakazad$^\star$, Thomas~B.~Sch\"on$^\dagger$, \\ Anders~Hansson$^\star$ and Jeroen~D.~Hol$^\ddagger$ \\
\small{$^\star$Department of Electrical Engineering, Link\"oping University, SE-581~83~Link\"oping, Sweden} \\
\small{E-mail: \{manon.kok, sina.khoshfetratpakazad, anders.g.hansson\}@liu.se} \\
\small{$^\dagger$Department of Information Technology, Uppsala University, Sweden} \\
\small{E-mail: thomas.schon@it.uu.se} \\
\small{$^\ddagger$Xsens Technologies B.V., Enschede, the Netherlands} \\
\small{E-mail: jeroen.hol@xsens.com} \\
}

\maketitle

\begin{abstract}
In inertial motion capture, a multitude of body segments are equipped with inertial sensors, consisting of 3D accelerometers and 3D gyroscopes. Using an optimization-based approach to solve the motion capture problem allows for natural inclusion of biomechanical constraints and for modeling the connection of the body segments at the joint locations. The computational complexity of solving this problem grows both with the length of the data set and with the number of sensors and body segments considered. In this work, we present a scalable and distributed solution to this problem using tailored message passing, capable of exploiting the structure that is inherent in the problem. As a proof-of-concept we apply our algorithm to data from a lower body configuration.
\end{abstract}

\section{Introduction}
\label{sec:introduction}
\begin{figure}[h]
\begin{center}
\includegraphics[width = 0.3\columnwidth]{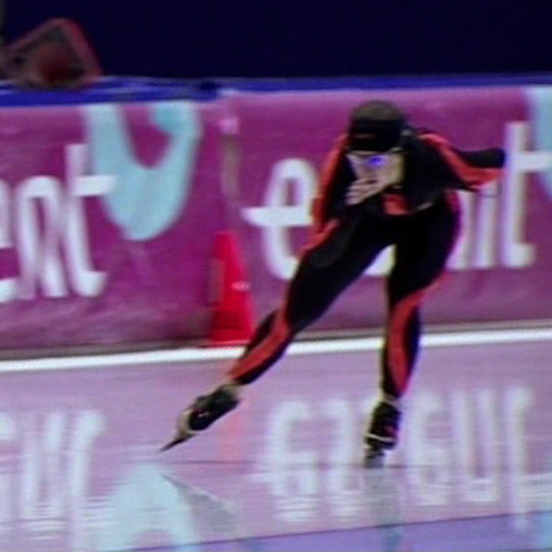}%
\includegraphics[width = 0.3\columnwidth]{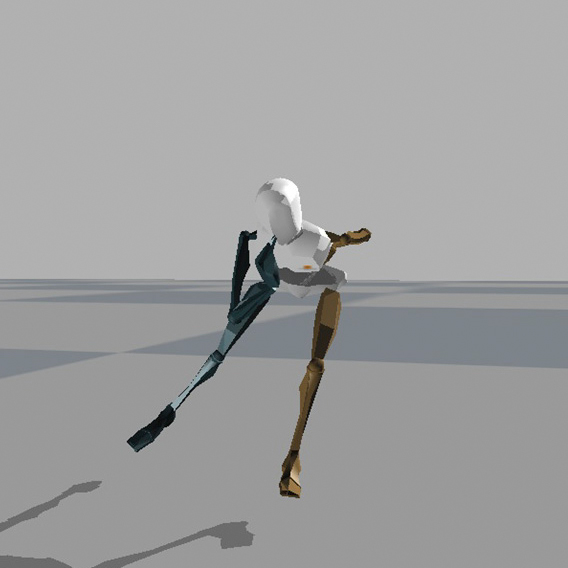} \\
\caption{\small Example of inertial motion capture. Left: olympic and world champion speed skating Ireen W\"ust wearing an inertial motion capture suit with $17$ inertial sensors. Right: graphical representation of the estimated position and orientation of the body segments. \normalsize}
\label{fig:inertialMotionCapture}
\end{center}
\end{figure}

Inertial motion capture focuses on estimating the relative position and orientation (pose) of different human body segments. To this end, inertial sensors (3D accelerometers and 3D gyroscopes) are placed on different body segments as shown in \Figureref{fig:inertialMotionCapture}. Each body segment's pose can be estimated by integrating the gyroscope data and double integrating the accelerometer data in time and combining these integrated estimates with a biomechanical model. Inertial sensors are successfully used for full body motion capture in many applications such as character animation, sports and biomechanical analysis~\cite{xsens,roetenbergLS:2013,kangJPK:2011,yunB:2006}.

In~\cite{kokHS:2014}, an optimization-based solution to the inertial motion capture problem was presented. It post-processes the data to obtain a smoothing estimate of the body's relative pose. The problem is solved using sequential quadratic programming (SQP)~\cite{nocedalW:2006}. The method was shown to result in drift-free and accurate pose estimates. Using an optimization-based approach allows for natural inclusion of biomechanical constraints and for modeling the connection between the body segment at the joint locations. Furthermore, it naturally handles nonlinearities and opens up the possibility for incorporating non-Gaussian noise and for simultaneous estimation of calibration parameters.

For applications which require real-time pose estimates, approximate solutions to the full smoothing problem need to be considered, for instance using filtering or moving horizon estimation (MHE)~\cite{raoRL:2001}. In these approaches, data up to a current time point is used to estimate the current pose. However, in case real-time estimates are not required, all available data can be used to obtain a smoothing estimate. Compared to filtering and MHE, obtaining a smoothing estimate is computationally more expensive and can be challenging both due to the computational complexity of solving the problem and due to storage requirements for constructing the problem. This is specifically of concern when processing long data sets.

In this paper we solve the same problem as in~\cite{kokHS:2014}. Again we use SQP, but at each iteration we compute the search directions using the message passing algorithm presented in~\cite{khoshfetratPakazadHA:2015}. This allows us to efficiently make use of the structure inherent in the problem. We exploit this structure in two different ways:
\begin{enumerate}
\item We reorder the problem based on time. This allows us to solve the problem by solving a large number of small problems which enables us to process long data sets.
\item We reorder the problem based on sensors and body segments. This leads to less computational benefits -- the number of sensors and body segments is typically much smaller than the number of time steps considered -- but it allows for solving the problem in a \emph{distributed} manner. It also relaxes the need for a centralized unit and streaming of data to it.
 \end{enumerate}
Using message passing for computing the search directions for the time-ordered problem has close connections to serial dynamic programming~\cite{bertsekas:1995}. This is due to the chain-like coupling structure in the problem. In fact, using serial dynamic programming, the search directions can be computed by sweeping through the available data forward and backwards, similar to the approach used for Rauch-Tung-Striebel (RTS) smoothing~\cite{rauchST:1965}. Using message passing, we compute the search directions by simultaneously starting from the first and final time  steps and sweeping towards the middle of the data set and back. This allows us to speed up the search direction computation by a factor of two. Notice that unlike existing scalable algorithms for solving big data problems that rely on first-order methods, see \eg \cite{cevherBS:2014}, the proposed algorithm solely relies on second-order methods. Consequently, this algorithm enjoys a far superior superlinear convergence rate,~\cite{wright:1997}, in comparison to at best linear convergence of other algorithms. 
 
If we only consider the lower body for the sensor-ordered problem, the chain-like coupling structure will also be present in the problem. Instead of running through time, this chain runs from one foot through both legs to the other foot. Consequently, it enjoys the same similarities to serial dynamic programming as discussed above. For the full body, the coupling structure will not be chain-like. It will, however, have an inherent tree structure. Hence, we can still use message passing for computing the search directions. In this paper, we focus on the lower body to simplify both the notation and the biomechanical modeling. The presented material can, however, straightforwardly be extended to the full body problem. 

The paper is organized as follows. In \Sectionref{sec:problemFormulation} we introduce the inertial motion capture problem for which the models are subsequently introduced in \Sectionref{sec:model}. In \Sectionref{sec:structureProblem}, we reorder the problem in the two ways described above. These two equivalent formulations of the original problem enjoy a special structure which allows us to use message passing to compute the search directions. The message passing algorithm will be introduced in \Sectionref{sec:messagePassing}. The resulting algorithm that can be used to solve the reordered problems is subsequently discussed in \Sectionref{sec:usingMPforMC}. In \Sectionref{sec:results}, we will discuss experimental results where the algorithm is applied to data from inertial sensors placed on the lower body. 

\section{Problem formulation}
\label{sec:problemFormulation}
The problem of estimating the relative pose of each body segment is formulated as a constrained estimation problem. Given $N_T$ measurements $y = \{y_1, \hdots, y_{N_T} \}$, a point estimate of the static parameters $\theta$ and the time-varying variables $x = \{ x_1, \hdots , x_{N_T} \}$ can be obtained as a constrained maximum a posteriori (MAP) estimate,
\begin{equation}
\begin{aligned}
\maximize_{x,\theta}& \quad p(x, \theta \mid y) \\
\st& \quad c( x ) = 0,
\label{eq:mapEstimate}
\end{aligned}
\end{equation}
where $c(x)$ represents the equality constraints. In this work we consider $N_S$ sensors placed on $N_S$ body segments, where sensor $\text{S}_i$ is placed on body segment $\text{B}_i$. The time-varying variables $x$ consist of variables both related to sensors (\eg the pose of the sensor) and to body segments (the pose of the body segment), \ie $x_t = \{ x^{\text{S}_1}_t, \hdots, x^{\text{S}_{N_S}}_t, x^{\text{B}_1}_t, \hdots, x^{\text{B}_{N_S}}_t \}$. To refer to the time-varying variables for sensor $\text{S}_i$, we use the notation $x^{\text{S}_i} = \{ x^{\text{S}_i}_{1}, \hdots, x^{\text{S}_{i}}_{N_T} \}$, while $x^{\text{B}_i} = \{ x^{\text{B}_i}_{1}, \hdots, x^{\text{B}_{i}}_{N_T} \}$ denotes time-varying variables for body segment $\text{B}_i$. The set of time-varying variables pertaining to sensor $\text{S}_i$ and segment $\text{B}_i$ is denoted $x^i = \{ x^{\text{B}_i}, x^{\text{S}_i}\}$. The static parameters are given by $\theta = \{ \theta^{\text{S}_1}, \hdots, \theta^{\text{S}_{N_S}} \}$. This notation will be used throughout this work and is summarized in \Tableref{table:notation}.

\begin{table*}
\caption{\small Notation to refer to the variables and the constraints in our problem, introduced in Sections~\ref{sec:problemFormulation} and~\ref{sec:bioconstraints}, respectively. \normalsize}
\begin{center}
\begin{tabular}{c r@{$\, = \,$} l m{0.3\textwidth}}
\toprule
Symbol & \multicolumn{2}{c}{Definition} & Explanation \\
\midrule
$x^{\text{S}_i}$ & $x^{\text{S}_i}$ & $\{ x^{\text{S}_i}_{1}, \hdots, x^{\text{S}_{i}}_{N_T} \}$& Time-varying variables pertaining to sensor $\text{S}_i$ \\
$x^{\text{B}_i}$ & $x^{\text{B}_i}$ & $\{ x^{\text{B}_i}_{1}, \hdots, x^{\text{B}_{i}}_{N_T} \}$& Time-varying variables pertaining to body segment $\text{B}_i$ \\
$x^i$ & $x^i$ & $\{ x^{\text{B}_i}, x^{\text{S}_{i}} \}$& Time-varying variables pertaining to sensor $\text{S}_i$ and body segment $\text{B}_i$\\
$x_t$ & $x_t$ & $\{ x^{\text{S}_1}_{t}, \hdots, x^{\text{S}_{N_S}}_{t}, x^{\text{B}_1}_{t}, \hdots, x^{\text{B}_{N_S}}_{t} \}$& Time-varying variables pertaining to time $t$ \\
$x$ & $x$ & $\{ x_1, \hdots, x_{N_T} \} $& All time-varying variables \\
$\theta$ & $\theta$ & $\{ \theta^{\text{S}_1}, \hdots, \theta^{N_S} \}$ & Static parameters \\
\midrule
$c^i(x^i,x^{i+1})$ & $c^i(x^i,x^{i+1})$ & $\{ c_1^i(x^i_1,x^{i+1}_1), \hdots, c_{N_T}^i(x^i_{N_T},x^{i+1}_{N_T}) \}$ & Biomechanical constraints for joint $i$ at time $t = 1, \hdots, N_T$ \\
$c_t(x_t)$ & $c_t(x_t)$ & $\{ c_t^1(x^1_t,x^2_t), \hdots, c_t^{N_S-1}(x_t^{N_S-1},x_t^{N_S}) \}$ & Biomechanical constraints at time $t$ \\
$c(x)$ & $c(x)$ & $\{ c_t(x_1), \hdots, c_t(x_{N_T}) \}$ & All biomechanical constraints \\
\bottomrule
\end{tabular}
\end{center}
\label{table:notation}
\end{table*}

Using the Markov property of the time-varying variables and the fact that the logarithm is a monotonic function, we can rewrite \eqref{eq:mapEstimate} as
\begin{equation}\label{eq:optimizationProblem}
\begin{split}
\minimize_{x,\theta} &
\underbrace{- \sum_{t = 2}^{N_T} \sum_{i = 1}^{N_S} \log p(x^{\text{S}_i}_t \mid x^{\text{S}_i}_{t-1}, \theta^{\text{S}_i}, y_t^{\text{S}_i})}_{\text{dynamics of the state }x^{\text{S}_i}_t}
\\ &
\underbrace{- \sum_{t = 1}^{N_T} \sum_{i = 1}^{N_S} \log p(x^{\text{B}_i}_t \mid x^{\text{S}_i}_{t})}_{\text{placement of sensor $\text{S}_i$ on body segment $\text{B}_i$}}\\
& \hspace{20mm}\underbrace{- \sum_{i = 1}^{N_S} \log p(x^{\text{S}_i}_1 \mid y_1^{\text{S}_i}) - \sum_{i = 1}^{N_S}\log p(\theta^{\text{S}_i})}_{\text{prior}}
\\
\st& \hspace{1mm} c(x) = 0.
\end{split}
\end{equation}
The constraints $c(x)$ represent the connection between the body segments at the joint locations. The cost function consists of terms related to a dynamic model for the time-varying states $x^{\text{S}_i}_t$, a model regarding the placement of the sensors on the body segments and a prior on the initial states $x^{\text{S}_i}_1$ and the constant parameters $\theta^{\text{S}_i}$ for $i = 1, \dots, N_S$.

\section{Model}
\label{sec:model}
To estimate the relative pose of the lower body, we assume that $7$ sensors are placed on different body segments. For notational simplicity, we assume that sensor $\text{S}_i$ is attached to body segment $\text{B}_i$. The body segments are connected at the joint locations. \Figureref{fig:coordinateSystem} illustrates two body segments, which can be thought of as the upper leg ($\text{B}_3$) and the lower leg ($\text{B}_2$). A sensor is attached to each body segment and the body segments are connected at the joint $\text{J}_2$ (the knee). Estimating the relative pose of the body amounts to estimating the position and orientation of the sensors and the body segments using the sensor measurements and the information that the body segments are connected. The variables considered optimization problem~\eqref{eq:optimizationProblem} are given by:
\begin{figure}
\begin{center}
\includegraphics[width = 0.3\columnwidth]{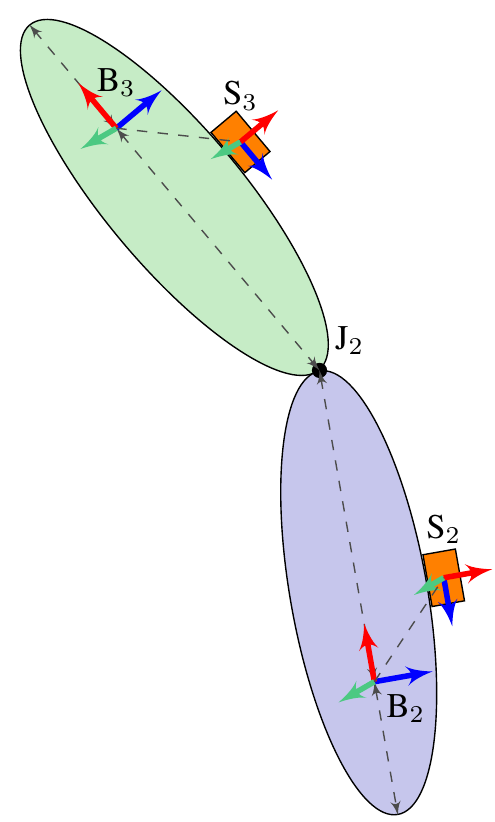}%
\end{center}
\caption{\small Connection of two body segments with a sensor attached to each of them.\normalsize}
\label{fig:coordinateSystem}
\end{figure}
\begin{table}[t]
\caption{\small Summary of the body segments, sensors and joints used in the model. \normalsize}
\begin{center}
\begin{tabularx}{0.78\linewidth}{l c l *{1}{C}}
\toprule
Body segment & Sensor & Joint & Connecting body segments \\
\cmidrule(lr){1-2} \cmidrule(lr){3-4}
$\text{B}_1$: Right foot & $\text{S}_1$ & $\text{J}_1$: Right ankle & $\text{B}_1 \Leftrightarrow \text{B}_2$ \\
$\text{B}_2$: Right lower leg & $\text{S}_2$ & $\text{J}_2$: Right knee & $\text{B}_2 \Leftrightarrow \text{B}_3$ \\
$\text{B}_3$: Right upper leg & $\text{S}_3$ & $\text{J}_3$: Right hip & $\text{B}_3 \Leftrightarrow \text{B}_4$ \\
$\text{B}_4$: Pelvis & $\text{S}_4$ & $\text{J}_4$: Left hip & $\text{B}_4 \Leftrightarrow \text{B}_5$ \\
$\text{B}_5$: Left upper leg & $\text{S}_5$ & $\text{J}_5$: Left knee & $\text{B}_5 \Leftrightarrow \text{B}_6$ \\
$\text{B}_6$: Left lower leg & $\text{S}_6$ & $\text{J}_6$: Left ankle & $\text{B}_6 \Leftrightarrow \text{B}_7$ \\
$\text{B}_7$: Left foot & $\text{S}_7$ \\
\bottomrule
\end{tabularx}
\end{center}
\label{table:biomechanicalModel}
\end{table}
\begin{itemize}
\item The time-varying variables $x_t^{\text{S}_i}$, consisting of the 3D position, velocity and orientation of sensor $\text{S}_i$ at time $t$. Furthermore, for one of the sensors $\text{S}_i$, the variables $x_t^{\text{S}_i}$ also include variables to estimate its mean acceleration at time $t$.
\item The time-varying variables $x_t^{\text{B}_i}$ consisting of the 3D position and orientation of body segment $\text{B}_i$ at time $t$.
\item The constant variables $\theta^{\text{S}_i}$ consisting of the gyroscope bias $b_{\omega}^{\text{S}_i} \in \mathbb{R}^3$ of sensor $\text{S}_i$.
\end{itemize}
Hence, the variables in the optimization problem are $x \in \mathbb{R}^{(15N_S + 3)N_T}$ and $\theta \in \mathbb{R}^{3N_S}$, where it is assumed that the orientation variables are encoded using a three-dimensional vector, see \eg\cite{crassidisMC:2007, grisettiKSFH:2010, hol:2011}. In the remainder of this section, we discuss the \textit{structure} of the cost functions and of the constraints in~\eqref{eq:optimizationProblem}. This structure will be exploited in our message passing algorithm. For an alternative and more explicit formulation of the problem, we refer to~\cite{kokHS:2014}.

\subsection{Dynamics of the state $x_t^{\text{S}_i}$}
The dynamics in~\eqref{eq:optimizationProblem} expresses the position, velocity and orientation of each sensor $\text{S}_i$ in terms of their values at the time instance $t-1$ and in terms of the constant variables $\theta^{\text{S}_i}$. The \textit{change} in position, velocity and orientation of sensor $\text{S}_i$ is modeled in terms of the acceleration and angular velocity measured by sensor $\text{S}_i$. The mean acceleration is modeled in terms of $x_{t-1}^{\text{S}_i}$, $\theta^{\text{S}_i}$ and the accelerometer measurements. For more details on the acceleration model we refer to~\cite{kokHS:2014}. The dynamics of the state $x_t^{\text{S}_i}$ can hence be expressed as in~\eqref{eq:optimizationProblem}.

\subsection{Placement of the sensors on the body segments}
As shown in \Figureref{fig:coordinateSystem}, sensor $\text{S}_i$ is assumed to be attached to the body segment $\text{B}_i$. We assume that the relative position and orientation of the sensors on the body segments is known from calibration. At each time instance, the position and orientation  of sensor $\text{S}_i$ can therefore be expressed in terms of the position and orientation of the body segment $\text{B}_{i}$. Ideally, this can be incorporated using equality constraints in~\eqref{eq:optimizationProblem}. However, it is physically impossible to place the sensor directly on the bone. Hence, it has to be placed on the soft tissue and the sensor will inevitably move slightly with respect to the bone. To allow for small random movements of the sensor, we incorporate the knowledge about the placement of the sensors on the body segments in the cost function.

\subsection{Biomechanical constraints}
\label{sec:bioconstraints}
The constraints $c(x)$ in the optimization problem~\eqref{eq:optimizationProblem} enforce the body segments to be connected at the joint locations at all times. Hence, for joint $\text{J}_i$, they model the position and the orientation of body segment $\text{B}_i$ in terms of the position and the orientation of body segment $\text{B}_{i+1}$ for $i = 1, \hdots, N_S - 1$. Here, the ordering of the indices of the joints and the body segments is assumed to be as in \Tableref{table:biomechanicalModel}. Note that we assume that the length of the body segments is known either from calibration or from a biomechanical model. 

Each joint $\text{J}_i$ results in a constraint $c^i_t \in \mathbb{R}^3$ at time $t$. The set of constraints at time $t$ is given by $c_t(x_t) = \{ c_t^1(x^1_t,x^2_t), \hdots, c_t^{N_S-1}(x_t^{N_S-1},x_t^{N_S}) \}$ and the set of constraints for joint $\text{J}_i$ is given by $c^i(x^i,x^{i+1}) = \{ c_1^i(x_1^i,x_1^{i+1}), \hdots, c_{N_T}^i(x_{N_T}^i,x_{N_T}^{i+1}) \}$. The complete set of biomechanical constraints is given by $c(x) = \{ c_t(x_1), \hdots, c_t(x_{N_T}) \}$. This notation is summarized in \Tableref{table:notation}. Note that we explicitly indicate which states are involved in the constraints using the ordering of body segments and joints in \Tableref{table:biomechanicalModel}.

\section{Problem Reformulation Enabling Structure Exploitation}
\label{sec:structureProblem}
In this section we focus on reordering the problem~\eqref{eq:optimizationProblem} in two different ways. In \Sectionref{sec:timeOrdering}, we reorder the problem based on the time indices $t = 1, \hdots, N_T$. In \Sectionref{sec:sensorOrdering}, we reorder the problem based on sensor and body segment indices $i = 1, \hdots, N_S$. The inertial motion capture problem can be solved iteratively using SQP, where at each iteration $k$ we solve a quadratic approximation of~\eqref{eq:optimizationProblem}. Hence, in each of the sections below, we also introduce an explicit formulation of the quadratic approximation that needs to be solved, where the reordering will allow us to exploit the structure inherent in the problem.

\subsection{Reordering based on time}
\label{sec:timeOrdering}
The objective function in~\eqref{eq:optimizationProblem} can be reordered based on time resulting in

\begin{equation}\label{eq:optimizationProblemTimeOrdered}
\begin{split}
\minimize_{x,\theta} &-
\sum_{i = 1}^{N_S}\left( \log p(x^{\text{B}_i}_1 \mid x^{\text{S}_i}_{1}) + \log p(x^{\text{S}_i}_1 \mid y_1^{\text{S}_i}) + \tfrac{1}{N_T} \log p(\theta^{\text{S}_i}) \right)  \\
&- \sum_{t = 2}^{N_T} \sum_{i = 1}^{N_S} \left( \log p(x^{\text{S}_i}_t \mid x^{\text{S}_i}_{t-1}, \theta^{\text{S}_i})+ \log p(x^{\text{B}_i}_t \mid x^{\text{S}_i}_{t}) + \tfrac{1}{N_T} \log p(\theta^{\text{S}_i})\right)
\\
\st& \hspace{1mm} c(x) = 0.
\end{split}
\end{equation}
Let $f_1(x_1, \theta)$ and $f_t(x_t,x_{t-1},\theta)$ for $t = 2, \dots, N_T$ correspond to different terms in the cost function of \eqref{eq:optimizationProblemTimeOrdered}. We can then rewrite \eqref{eq:optimizationProblemTimeOrdered} more compactly as
\begin{equation} \label{eq:optimizationProblemTimeOrderedIntermed}
\begin{split}
\minimize_{x, \theta} & \quad f_1(x_1,\theta) + \sum_{t=2}^{N_T} f_t(x_t, x_{t-1},\theta)\\
\st & \quad  c_t(x_t) = 0, \quad t = 1, \dots, N_T,
\end{split}
\end{equation}
where we use the notation $c_t(x_t)$ to denote the biomechanical constraints at time $t$ as introduced in \Tableref{table:notation}. It is beneficial to equivalently reformulate this problem as
\begin{equation}\label{eq:optimizationProblemTimeOrderedReform}
\begin{split}
\minimize_{x, \bar \theta} & \quad f_1(x_1,\bar \theta_1) + \sum_{t=2}^{N_T} f_t(x_t, x_{t-1},\bar \theta_t)\\
\st & \quad c_t(x_t) = 0, \quad t = 1, \dots, N_T,\\
& \quad \bar \theta_{t} = \bar\theta_{t+1}, \quad t = 1, \dots, N_T-1,
\end{split}
\end{equation}
where $\bar \theta = \{ \bar \theta_1, \dots, \bar \theta_{N_T} \}$. This formulation models the \emph{constant} variables $\theta$ in terms of \emph{time-varying} variables $\bar \theta_t$. Inclusion of the additional equality constraints in~\eqref{eq:optimizationProblemTimeOrderedReform} ensures that $\bar \theta_t$ will be equal for all $t$ and makes the formulations~\eqref{eq:optimizationProblemTimeOrderedIntermed} and~\eqref{eq:optimizationProblemTimeOrderedReform} equivalent. 

\begin{figure}
\begin{center}
\includegraphics[width = 0.4\columnwidth]{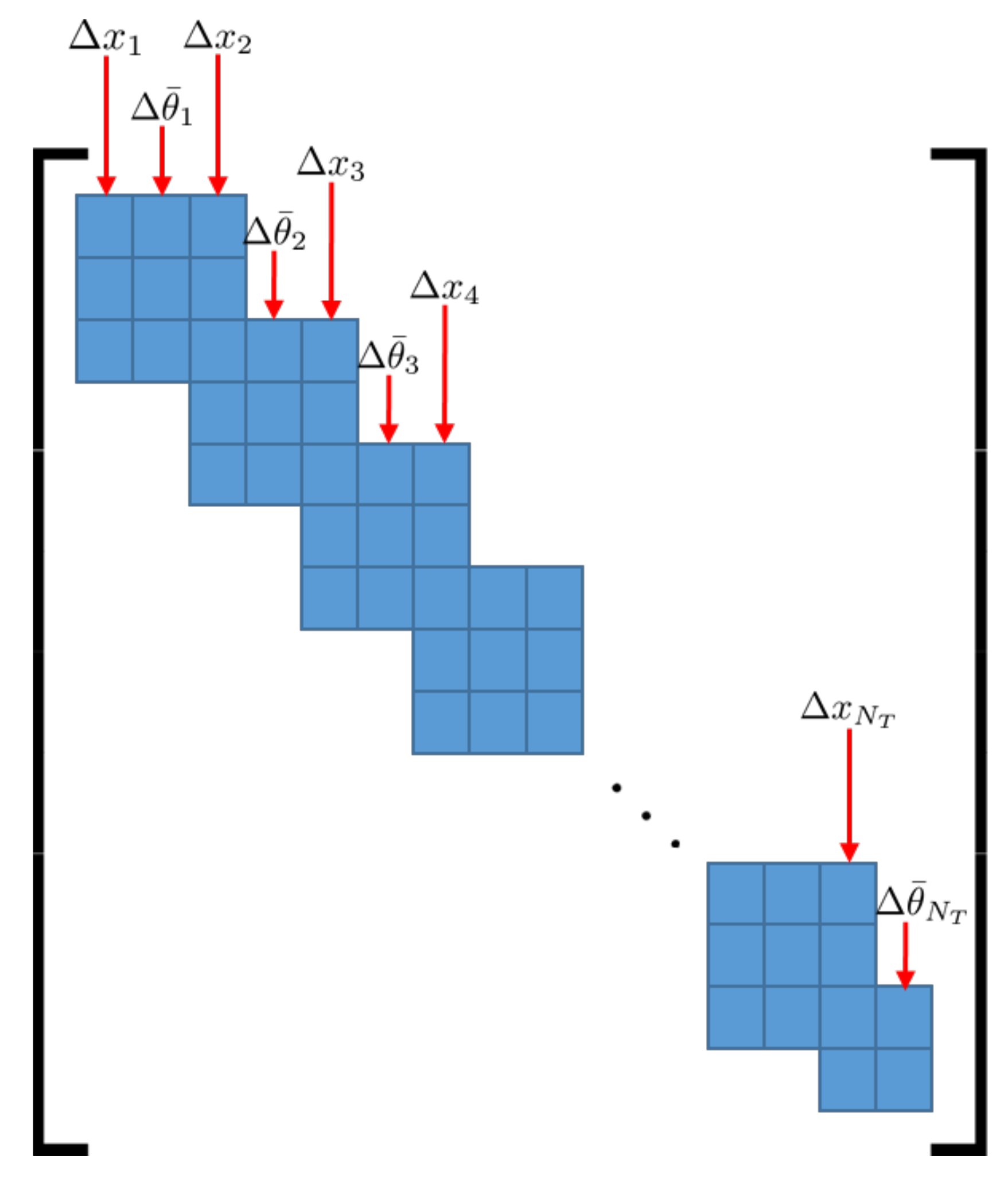}
\end{center}
\caption{\small Form of the Hessian $H(x,\bar \theta)$ for the quadratic approximation~\eqref{eq:optimizationProblemTimeOrderedReformQP} with the time-ordered variables as described in \Sectionref{sec:timeOrdering}. The blue blocks indicate the non-zero terms in the Hessian. For clarity, we indicate which variables are associated with which blocks. \normalsize}
\label{fig:hessianOrderTime}
\end{figure}

The reordered problem \eqref{eq:optimizationProblemTimeOrderedReform} enjoys a desirable structure that can be exploited. It can be solved iteratively using SQP, where at each iteration $k$ we solve the quadratic approximation

\begin{equation}\label{eq:optimizationProblemTimeOrderedReformQP}
\begin{split}
\minimize_{\Delta x, \Delta \bar \theta}& \quad \hspace{-2mm} \tfrac{1}{2}\begin{bmatrix}\Delta x \\ \Delta \bar  \theta \end{bmatrix}^\Transp H(x^{(k)},\bar \theta^{(k)})\begin{bmatrix}\Delta x \\ \Delta \bar  \theta \end{bmatrix} + \left( J_f(x^{(k)},\bar \theta^{(k)}) \right)^\Transp \begin{bmatrix}\Delta x \\ \Delta \bar  \theta \end{bmatrix} \\
\st& \quad \hspace{-2mm} c_t(x_t^{(k)}) + \left( J_{c_t}(x_t^{(k)}) \right)^\Transp \Delta x_t = 0, \ t = 1, \dots, N_T,\\
& \quad \hspace{-2mm} \Delta \bar \theta_t - \Delta \bar \theta_{t+1} = 0, \quad t = 1, \dots, N_T-1,
\end{split}
\end{equation}
to compute a step, $\begin{bmatrix}\Delta x^\Transp & \Delta \bar  \theta^\Transp \end{bmatrix}^\Transp$. This step will be used to update the estimates of the variables $x$ and $\bar \theta$. The Jacobians of the objective function and of the constraints are given by
\begin{subequations}
\begin{align}
J_f(x,\bar \theta) &= \nabla_{x,\bar \theta} f_1(x_1,\bar \theta_1) + \sum_{t=2}^{N_T} \nabla_{x,\bar \theta} f_t(x_t, x_{t-1},\bar \theta_t), \\
J_{c_t}(x_t) &= \nabla_{x_t} c_t(x_{t}).
\end{align}
\end{subequations}
For the Hessian of the objective function we use a Gauss-Newton approximation as
\begin{align}
H(x, \bar \theta) \approx   \nabla_{x,\bar \theta} f_1(x_1,\theta_1)\nabla_{x,\bar \theta} f_1(x_1,\bar \theta_1)^\Transp + \sum_{t=2}^{N_T} \nabla_{x,\bar \theta} f_t(x_t, x_{t-1},\bar \theta_t)\nabla_{x,\bar \theta} f_t(x_t, x_{t-1},\bar \theta_t)^\Transp.
\end{align}
If we choose the ordering of variables as $(\Delta x_1, \Delta \bar \theta_1, \Delta x_2, \Delta\bar \theta_2, \dots, \Delta x_{N_T}, \Delta\bar \theta_{N_T})$, the Hessian $H(x,\bar \theta)$ takes a special form as illustrated in \Figureref{fig:hessianOrderTime}. In this case it is possible to find matrices $H_t$ and $h_t$ and write the problem in \eqref{eq:optimizationProblemTimeOrderedReformQP} equivalently as

\begin{equation}\label{eq:optimizationProblemTimeOrderedReformQPReorder}
\begin{split}
\minimize_{\Delta x, \Delta \bar \theta}& \quad  \sum_{t = 1}^{N_T-1}\left( \tfrac{1}{2}\begin{bmatrix}\Delta x_t \\ \Delta \bar \theta_{t}\\ \Delta x_{t+1} \\ \Delta\bar\theta_{t+1} \end{bmatrix}^\Transp H_t \begin{bmatrix}\Delta x_t \\ \Delta\bar\theta_{t}\\ \Delta x_{t+1} \\ \Delta\bar\theta_{t+1} \end{bmatrix} +  \begin{bmatrix}\Delta x_t \\ \Delta\bar\theta_{t}\\ \Delta x_{t+1} \\ \Delta\bar\theta_{t+1} \end{bmatrix}^\Transp h_t \right) \\
\st& \quad c_t(x_t^{(k)}) + \left( J_{c_t}(x_t^{(k)}) \right)^\Transp \Delta x_t = 0, \ t = 1, \dots, N_T,\\
& \quad \Delta \bar \theta_t - \Delta \bar \theta_{t+1} = 0, \quad t = 1, \dots, N_T-1.
\end{split}
\end{equation}
This time-ordered equivalent formulation of the problem~\eqref{eq:optimizationProblem} enjoys a special structure which allows us solve it efficiently using message passing. Before introducing this approach, we will first reorder the problem~\eqref{eq:optimizationProblem} in a second way, based on sensors and body segments.

\subsection{Reordering based on sensors and body segments}
\label{sec:sensorOrdering}
The problem~\eqref{eq:optimizationProblem} can also be rearranged or reordered based on sensors and body segments. Here, we group the terms in the cost function related to sensor $\text{S}_i$ and body segment $\text{B}_i$ for $i = 1, \hdots, N_S$, resulting in

\begin{equation}\label{eq:optimizationProblemSensoredOrdered}
\begin{split}
\minimize_{x,\theta}
& - \sum_{i = 1}^{N_S} \left(\log p(x^{\text{S}_i}_1 \mid y_1^{\text{S}_i}) + \sum_{t = 2}^{N_T} \log p(x^{\text{S}_i}_t \mid x^{\text{S}_i}_{t-1}, \theta^{\text{S}_i}) 
+ \sum_{t = 1}^{N_T} \log p(x^{\text{B}_i}_t \mid x^{\text{S}_i}_{t}) + \log p(\theta^{\text{S}_i})\right)
\\
\st& \hspace{1mm} c(x) = 0.
\end{split}
\end{equation}
Letting each term in the cost function be denoted by $g^i(x^i,\theta^{\text{S}_i})$, we can write~\eqref{eq:optimizationProblemSensoredOrdered} compactly as
\begin{equation}\label{eq:optimizationProblemSensorOrderedCompact}
\begin{split}
\minimize_{x, \theta} & \quad \sum_{i=1}^{N_S} g^i(x^i, \theta^{\text{S}_i})\\
\st & \quad  c^i(x^i,x^{i+1}) = 0, \quad i = 1, \dots, N_S-1,
\end{split}
\end{equation}
where the constraints are grouped per joint. Note again that $x^i$ and $c^i(x^i,x^{i+1})$ are defined in \Tableref{table:notation}.

Analogously to the development in \Sectionref{sec:timeOrdering}, solving the problem in \eqref{eq:optimizationProblemSensorOrderedCompact} using SQP amounts to solving
\begin{equation}\label{eq:optimizationProblemSensorOrderedCompactQP}
\begin{split}
\minimize_{\Delta x, \Delta  \theta} & \quad \tfrac{1}{2}\begin{bmatrix}\Delta x \\ \Delta  \theta \end{bmatrix}^\Transp \bar H(x^{(k)},\theta^{(k)})\begin{bmatrix}\Delta x \\ \Delta  \theta \end{bmatrix} + \left( J_g(x^{(k)},\theta^{(k)}) \right)^\Transp \begin{bmatrix}\Delta x \\ \Delta \bar  \theta \end{bmatrix} \\
\st&  \quad c^i(x^{i,(k)},x^{i+1,(k)}) + \left( J_{c^i}(x^{i,(k)},x^{i+1,(k)}) \right)^\Transp \begin{bmatrix} \Delta x^i \Delta x^{i+1} \end{bmatrix} = 0,  \quad i = 1, \dots, N_S-1,
\end{split}
\end{equation}
at each iteration, where
\begin{subequations}
\begin{align}
J_g(x^{(k)},\theta^{(k)}) &= \sum_{i=1}^{N_S} \nabla_{x,\theta} g^i(x^{i}, \theta^{\text{S}_i}), \\
J_{c^i}(x^{i},x^{i+1}) &= \nabla_{x^{i},x^{i+1}} c^i(x^{i},x^{i+1}).
\end{align}
\end{subequations}
The Hessian of the objective function of this problem is again based on a Gauss-Newton approximation,
\begin{align}
\bar H(x,\theta) \approx  \sum_{i=1}^{N_S} \nabla_{x,\theta} g^i(x^{i}, \theta^{\text{S}_i})\nabla_{x,\theta} g^i(x^i, \theta^{\text{S}_i})^\Transp.
\end{align}
If we choose the ordering of variables as $(x^1, \theta^{\text{S}_1},x^{2}, \theta^{\text{S}_2}, \dots, x^{N_S}, \theta^{\text{S}_{N_S}})$, the Hessian becomes block-diagonal with each block corresponding to sensor $\text{S}_i$ and body segment $\text{B}_i$. This then enables us to write the problem in \eqref{eq:optimizationProblemSensorOrderedCompactQP} as
\begin{equation}\label{eq:optimizationProblemTimeOrderedCompactQPReorder}
\begin{split}
\minimize_{\Delta x, \Delta  \theta}& \  \sum_{i = 1}^{N_S-1}\left( \tfrac{1}{2}\begin{bmatrix}\Delta x^i \\ \Delta \theta^{\text{S}_i}\\ \Delta x^{i+1} \\ \Delta \theta^{\text{S}_{i+1}} \end{bmatrix}^\Transp \bar H^i\begin{bmatrix}\Delta x^{i} \\ \Delta \theta^{\text{S}_i}\\ \Delta x^{i+1} \\ \Delta \theta^{S_{i+1}} \end{bmatrix} +  \begin{bmatrix}\Delta x^{i} \\ \Delta \theta^{\text{S}_i}\\ \Delta x^{i+1} \\ \Delta \theta^{\text{S}_{i+1}} \end{bmatrix}^\Transp \bar h^i\right) \\
\st& \quad c^i(x^{i,(k)},x^{i+1,(k)}) + \left( J_{c^i}(x^{i,(k)},x^{i+1,(k)}) \right)^\Transp \begin{bmatrix} \Delta x^i \\ \Delta x^{i+1} \end{bmatrix} = 0, \quad i = 1, \dots, N_S-1,
\end{split}
\end{equation}
through consistent choices of matrices $\bar H^i$ and vectors $\bar h^i$. The problem formulation~\eqref{eq:optimizationProblemTimeOrderedCompactQPReorder} again enjoys a special structure which allows us to solve it efficiently using message passing. Next we briefly review this approach.

\section{Tree Structure in Coupled Problems and Message Passing}\label{sec:TreeStructureMP}
\label{sec:messagePassing}
Consider the following coupled optimization problem
\begin{align}\label{eq:coupledProblem}
\minimize_z \quad f_1(z) + f_2(z) + \dots + f_{\nCliques}(z),
\end{align}
where $z \in \mathbb R^{n_z}$ and $f_\indexOneClique : \mathbb R^{n_z} \rightarrow \mathbb R$ for $\indexOneClique = 1, \dots, \nCliques$. This problem can be seen as a combination of $\nCliques$ subproblems, each of which is defined by a term in the cost function and depends only on a few elements of $z$. Note that $f_\indexOneClique$ can include indicator functions on constraints. Hence, the problem formulations of the inertial motion capture problem~\eqref{eq:optimizationProblemTimeOrderedReform}, \eqref{eq:optimizationProblemTimeOrderedReformQPReorder} for the time ordering and~\eqref{eq:optimizationProblemSensorOrderedCompact}, \eqref{eq:optimizationProblemTimeOrderedCompactQPReorder} for the sensor and body segment ordering, are of the form~\eqref{eq:coupledProblem}.

Let us denote the ordered set of indices of $z$ that each subproblem $\indexOneClique$ depends on by $C_\indexOneClique$. We can then equivalently rewrite~\eqref{eq:coupledProblem} as
\begin{align}\label{eq:CPS}
\minimize_{z} & \quad   \bar f_1\left(z_{_{C_1}}\right) + \dots + \bar f_{\nCliques}(z_{_{\nCliques}}),
\end{align}
where $z_{_{C_a}}$ is a $|C_a|$-dimensional vector that contains the elements of $z$ indexed by $C_a$, with $|C_a|$ denoting the number of elements in the set $C_a$. Also  the functions $\bar f_\indexOneClique \ : \ \mathbb R^{|C_\indexOneClique|} \rightarrow \mathbb R$ are lower dimensional descriptions of $f_\indexOneClique$s such that $f_\indexOneClique(z) = \bar f_\indexOneClique(z_{_{C_a}})$ for all $z$ and $\indexOneClique = 1, \dots, \nCliques$. It is possible to describe the coupling structure of the problem graphically using undirected graphs. Particularly, let us define the \emph{sparsity graph} of the problem as a graph $G_s(V_s,\mathcal E_s)$ with the vertex set $V_s = \left\{ 1, \dots, n_z \right\}$ and $(a,b) \in \mathcal E_s$ if and only if variables $z_a$ and $z_b$ appear in the same subproblem. Let us assume that each $C_a$ for $a = 1,  \dots, N_C$, be a \emph{clique} of this graph, where a clique is a maximal subset of $V_s$ that induces a complete subgraph on $G_s$. This in turn means that no clique is contained in another clique \cite{blp:94}. 
Assume furthermore that there exists a tree defined on $\mathbf C_{G_s}$ such that for every $C_\indexOneClique, C_\indexTwoClique \in\mathbf C_{G_s}$ where $\indexOneClique \neq \indexTwoClique$, $C_\indexOneClique \cap C_\indexTwoClique$ is contained in all the cliques in the path connecting the two cliques in the tree. This property is called the clique intersection property \cite{blp:94}. Graphs with this property have an \emph{inherent tree structure} and can be represented using a \emph{clique tree}.

Let us assume that the sparsity graph of the problem~\eqref{eq:CPS} has an inherent tree structure. The problem can then be solved distributedly using a message passing algorithm that utilizes the clique tree as its computational graph. This means that the nodes $V_\cliqueTreeIndex = \{ 1, \dots, \nCliques \}$ act as computational agents that communicate or collaborate with their neighbors defined by the edge set $\mathcal E_\cliqueTreeIndex$. The message-passing algorithm solves \eqref{eq:CPS} by performing an upward-downward pass through the clique tree, see e.g., \cite{khoshfetratPakazadHA:2015, kol:09} and references therein. The upward pass starts from the agents at the leaves of the tree, i.e., all agents $\indexOneClique \in \leaves(T)$, where every such agent computes and communicates the message
\begin{align}\label{eq:leavesmij}
m_{\indexOneClique \parent(\indexOneClique)}\left(z_{_{\separatorSet_{\indexOneClique \parent(\indexOneClique)}}}\right) = \minimum_{z_{_{C_\indexOneClique \setminus \separatorSet_{\indexOneClique\parent(\indexOneClique)}}}} \left\{ \bar f_\indexOneClique \left(z_{_{C_\indexOneClique}}\right) \right\},
\end{align}
to its parent, denoted by $\parent(\indexOneClique)$. Here $\separatorSet_{\indexOneClique \indexTwoClique} = C_\indexOneClique \cap C_\indexTwoClique$ is the so-called separator set of agents $\indexOneClique$ and $\indexTwoClique$. Then every agent $\indexOneClique$ that has received all messages from its children, computes and communicates the message
\begin{align}\label{eq:mij}
m_{\indexOneClique \parent(\indexOneClique)}\left(z_{_{\separatorSet_{\indexOneClique \parent(\indexOneClique)}}}\right) = \minimum_{z_{_{C_\indexOneClique \setminus \separatorSet_{\indexOneClique \parent(\indexOneClique)}}}} \left\{  \bar f_\indexOneClique \left(z_{_{C_\indexOneClique}}\right) +  \sum_{\indexThreeClique \in \children(\indexOneClique)} m_{\indexThreeClique \indexOneClique}\left(z_{_{\separatorSet_{\indexThreeClique \indexOneClique}}}\right) \right\},
\end{align}
with $\children(\indexOneClique)$ denoting the children of agent $\indexOneClique$, to its parent. This procedure is continued until we reach the agent, $r$, at the root. At this point, agent $r$ computes its corresponding optimal solution by solving
\begin{align}\label{eq:RLocalProblem}
z^\ast_{_{C_r}} = \argmin_{z_{_{C_r}}} \left\{  \bar f_r\left(z_{_{C_r}}\right)  + \sum_{\indexThreeClique \in \children(r)} m_{\indexThreeClique r}\left(z_{_{\separatorSet_{\indexThreeClique r}}}\right) \right\},
\end{align}
and initiates the downward pass by communicating this solution to its children. During the downward pass each agent $\indexOneClique$ having received the optimal solution $\left( z_{_{\separatorSet_{\parent(\indexOneClique) \indexOneClique}}}^\ast \right)^{\parent(\indexOneClique)}$ from its parent computes its corresponding optimal solution as 
\begin{align}\label{eq:LocalProblemGen}
z^\ast_{_{C_\indexOneClique}} = \argmin_{z_{_{C_\indexOneClique}}} \Bigg \{  \bar f_\indexOneClique \left(z_{_{C_\indexOneClique }}\right)  + \sum_{\indexThreeClique \in \children(\indexOneClique)} m_{\indexThreeClique \indexOneClique}\left(z_{_{\separatorSet_{\indexThreeClique \indexOneClique}}}\right) + \frac{1}{2} \left\| z_{_{\separatorSet_{\parent(\indexOneClique) \indexOneClique}}} - \left( z_{_{\separatorSet_{\parent(\indexOneClique) \indexOneClique}}}^\ast \right)^{\parent(\indexOneClique)}\right\|^2 \Bigg\},
\end{align}
and communicates this solution to its children. Once the downward pass is accomplished, all agents have computed their respective optimal solution and the algorithm is terminated. We have summarized this scheme in Algorithm \ref{alg:MP}.
\begin{algorithm}[t]
\caption{Distributed Optimization Using Message Passing}\label{alg:MP}
\begin{algorithmic}[1]
\small
\State{Given a sparsity graph $G_s$ of an optimization problem}
\State{\quad extract its cliques and a clique tree over the cliques;}
\State{\quad assign each subproblem to one and only one of the agents.}
\State{Set $\textrm{agents} = \{ 1, \dots, \nCliques \} \setminus {r}$ and $\textrm{elim} = \{ \}$.}
\State{Perform the upward pass as}
\While{$|\textrm{agents}| \neq 0$}
\For{$i \in \textrm{agents}$}
\If{$\children(\indexOneClique) \subseteq \textrm{elim}$}
\State\parbox[t]{\dimexpr\linewidth-\algorithmicindent-3em \relax}{%
    \setlength{\hangindent}{\algorithmicindent}%
This agent computes the message in \eqref{eq:mij} and communicates it to agent $\parent(\indexOneClique)$.}
\State{$\textrm{elim} = \textrm{elim} \cup \{\indexOneClique\}$.}
\EndIf
\EndFor
\State{$\textrm{agents} = \textrm{agents} \setminus \textrm{elim}$.}
\EndWhile
\State{Set $\textrm{agents} = \{ 1, \dots, \nCliques \}$ and $\textrm{elim} = \{ \} $.}
\State{Perform the downward pass as}
\While{$|\textrm{agents}| \neq 0$}
\For{$\indexOneClique \in \textrm{agents}$}
\If{$\parent(\indexOneClique) \subseteq \textrm{elim}$}
\State \parbox[t]{\dimexpr\linewidth-\algorithmicindent-3em \relax}{%
    \setlength{\hangindent}{\algorithmicindent}%
    This agent computes optimal solution as in \eqref{eq:LocalProblemGen} and com\-mu\-ni\-cates it to agents $\children(\indexOneClique)$.}
\State{$\textrm{elim} = \textrm{elim} \cup \{\indexOneClique\}$.}
\EndIf
\EndFor
\State{$\textrm{agents} = \textrm{agents} \setminus \textrm{elim}$.}
\EndWhile
\State By the end of the downward pass all agents have computed their optimal solutions and the algorithm is terminated.
\end{algorithmic}
\end{algorithm}
\begin{rem}
Notice that within the upward pass all agents that have received messages from their children can compute their messages simultaneously and in parallel. This also holds for the downward pass, as all agents that have received the optimal solution from their parents can compute their respective optimal solution in parallel. 
\end{rem}

\section{Scalable and Distributed Solutions Using Message Passing}
\label{sec:usingMPforMC}
\begin{figure}
\begin{center}
\includegraphics[width = 0.6\columnwidth]{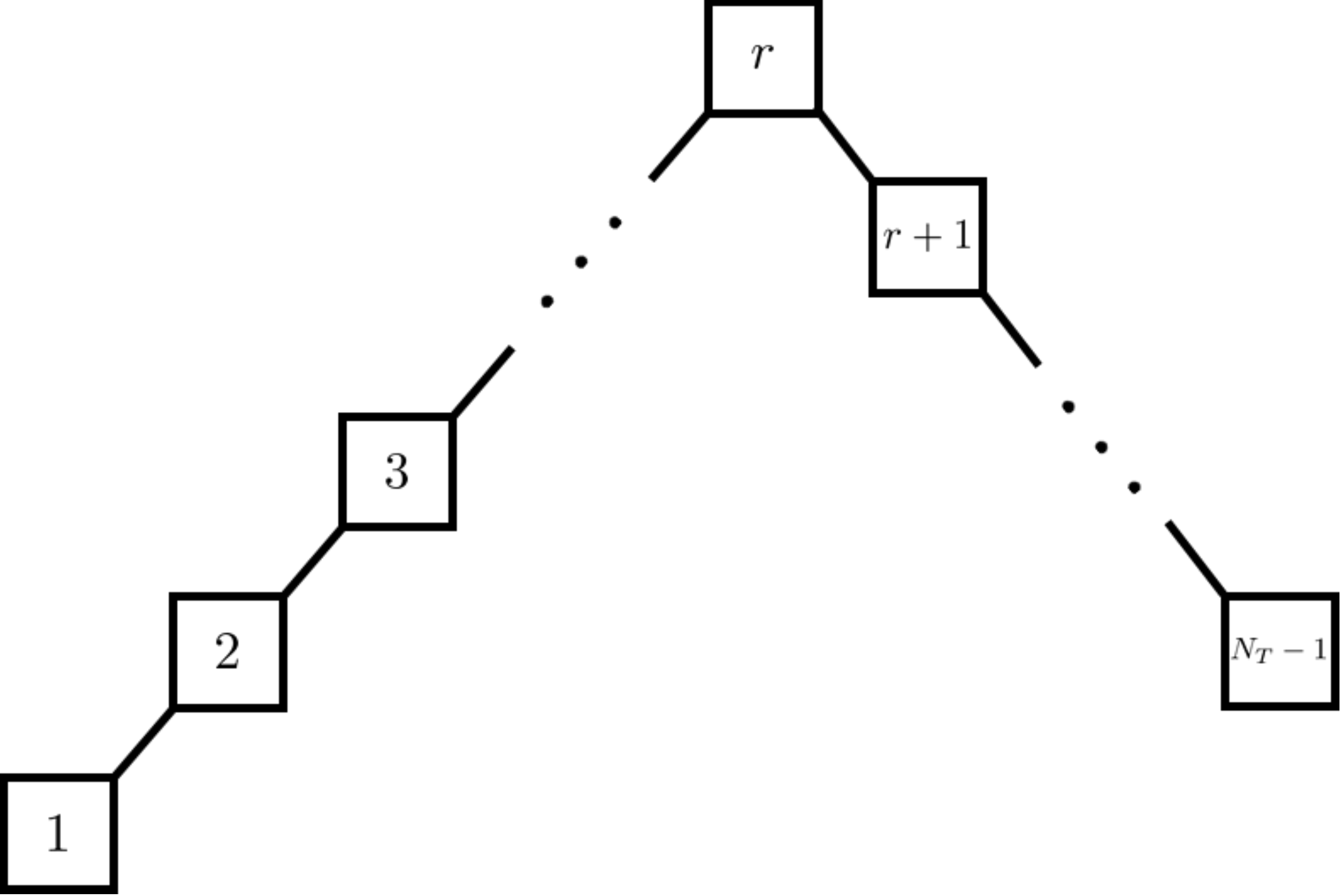}
\end{center}
\caption{Clique tree corresponding to the sparsity graph of problem \eqref{eq:optimizationProblemTimeOrderedReformQPReorder}.}
\label{fig:timetree}
\end{figure}
We will now rewrite the problem reorderings \eqref{eq:optimizationProblemTimeOrderedReformQPReorder} and \eqref{eq:optimizationProblemTimeOrderedCompactQPReorder} such that \Algorithmref{alg:MP} can be used to solve the problem. Let us first reconsider the problem in \eqref{eq:optimizationProblemTimeOrderedReformQPReorder}. We can rewrite this problem compactly as

\begin{subequations}\label{eq:optimizationProblemTimeOrderedReformQPReorderRegroup}
\begin{align}
\minimize_{\Delta x, \Delta \bar \theta}& \quad \sum_{t = 1}^{N_T-1} \bar f_t(\Delta x_t, \Delta x_{t+1},\Delta \bar \theta_{t},\Delta \bar \theta_{t+1})\\
\st& \notag\\& \begin{rcases*} c_t(x_t^{(k)}) + \left( J_{c_t}(x_t^{(k)}) \right)^\Transp \Delta x_t = 0 \\ \Delta \bar \theta_t - \Delta \bar \theta_{t+1} = 0 \end{rcases*}, t = 1, \dots, r-1,\\
& \begin{rcases*} c_t(x_t^{(k)}) + \left( J_{c_t}(x_t^{(k)}) \right)^\Transp \Delta x_t = 0 \\ \Delta \bar \theta_t - \Delta \bar \theta_{t+1} = 0 \\ c_t(x_{t+1}^{(k)}) + \left( J_{c_{t+1}}(x_{t+1}^{(k)}) \right)^\Transp \Delta x_{t+1} = 0\end{rcases*}, t = r,\\
& \begin{rcases*} c_t(x_{t+1}^{(k)}) + \left( J_{c_{t+1}}(x_{t+1}^{(k)}) \right)^\Transp \Delta x_{t+1} = 0 \\ \Delta \bar \theta_t - \Delta \bar \theta_{t+1} = 0 \end{rcases*}, t = r+1, \dots, N_T-1,
\end{align}
\end{subequations}
where $r = \lfloor N_T/2 \rfloor$. The sparsity graph of this problem has an inherent tree structure, with $N_T-1$ cliques. Each clique $C_\indexOneClique$ consists of the variables $\Delta x_\indexOneClique$, $\Delta \bar \theta_\indexOneClique$, $\Delta x_{\indexOneClique+1}$ and $\Delta \bar \theta_{\indexOneClique+1}$. The clique tree for this problem is illustrated in \Figureref{fig:timetree}. Consequently, we can use \Algorithmref{alg:MP} for solving this problem. During the upward pass, each agent $\indexOneClique$ sends a message as in~\eqref{eq:leavesmij} and~\eqref{eq:mij} to its parent as a function of the variables it shares with its parents. Hence, if $a < r$ (the agent is on the left side of agent $r$ in \Figureref{fig:timetree}) the message to its parents is a function of $\Delta x_{\indexOneClique+1}$ and $\Delta \bar \theta_{\indexOneClique+1}$. Equivalently, if $a > r$ (the agent is on the right side of agent $r$ in \Figureref{fig:timetree}) the message to its parents is a function of $\Delta x_{\indexOneClique}$ and $\Delta \bar \theta_{\indexOneClique}$. As a result, each agent except agent $r$ has to factorize a matrix of size $|x_\indexOneClique| + |\bar \theta_\indexOneClique|$ plus the number of constraints, which is equal to $6 N_S - 3$, as can be seen in~\eqref{eq:optimizationProblemTimeOrderedReformQPReorderRegroup}. The root agent instead has to factorize a matrix of size $2|x_\indexOneClique| + 2 |\bar \theta_\indexOneClique| + 9 N_S - 6$. The computational complexity of \Algorithmref{alg:MP} is dominated by the upward pass since the downward pass does not require a matrix factorization. For details on this, we refer to \cite{khoshfetratPakazadHA:2015}. Hence, the computational complexity and storage requirements for the resulting algorithm grow linearly with $N_T$. The reduction in the memory requirements follows from the fact that using Algorithm \ref{alg:MP} we have relaxed the need for even forming the problem in~\eqref{eq:optimizationProblemTimeOrderedReformQPReorder}. The resulting algorithm to solve the problem~\eqref{eq:optimizationProblemTimeOrderedReformQPReorderRegroup} is summarized in \Algorithmref{alg:resultingAlgorithm}.

The problem in \eqref{eq:optimizationProblemTimeOrderedCompactQPReorder} is also a coupled problem but with $N_S-1$ subproblems. The clique tree for this problem has the same structure as for \eqref{eq:optimizationProblemTimeOrderedCompactQPReorder}, where the only differences are in the number of cliques which in this case is $N_S-1$ and that $r = \lfloor N_S/2 \rfloor$. Each clique $C_\indexOneClique$ consists of the variables $\Delta x^\indexOneClique$, $\Delta \theta^{\text{S}_\indexOneClique}$, $\Delta x^{\indexOneClique+1}$ and $\Delta \theta^{\text{S}_{\indexOneClique+1}}$. Hence, we can solve the problem in \eqref{eq:optimizationProblemTimeOrderedCompactQPReorder} distributedly using \Algorithmref{alg:MP}. This can be achieved using a network of computational agents, that can be installed on the body and that collaborate based on the clique tree.

\begin{rem}
Note that in \eqref{eq:optimizationProblemTimeOrderedReformQPReorderRegroup} we have adopted a particular grouping of the constraints. This is to ensure that each of the subproblems is well-posed in terms of their local variables. This was not necessary for the problem in \eqref{eq:optimizationProblemTimeOrderedCompactQPReorder}.
\end{rem}
\begin{rem}
The reason that the clique trees for both problems in \eqref{eq:optimizationProblemTimeOrderedReformQPReorderRegroup} and \eqref{eq:optimizationProblemTimeOrderedCompactQPReorder} have the same structure is due to the fact that we have focused on the motion capture problem for the lower body. For solving the full body problem we can use the same approach as presented in this paper, since the inherent tree structure will still be present in the problem. However, the clique tree for the corresponding problem will be more complicated than a chain and will correspond to the body formation.
\end{rem}

\begin{algorithm}[t]
\caption{Inertial Motion Capture}\label{alg:resultingAlgorithm}
\begin{algorithmic}[1]
\small
\State Place the sensors on the body, calibrate the system and collect inertial measurements.
\State{Initialize $x$ and $\theta$ or $\bar \theta$ and set $k = 0$.}
\While{the algorithm has not converged and the solution is not feasible}
\State \parbox[t]{\dimexpr\linewidth-\algorithmicindent\relax}{%
    \setlength{\hangindent}{\algorithmicindent}%
    Formulate the quadratic approximation~\eqref{eq:optimizationProblemTimeOrderedReformQPReorderRegroup} using the time reordering or~\eqref{eq:optimizationProblemTimeOrderedCompactQPReorder} using the sensor / segment reordering.} \label{step:formProblem}
\State \parbox[t]{\dimexpr\linewidth-\algorithmicindent\relax}{%
    \setlength{\hangindent}{\algorithmicindent}%
    Use \Algorithmref{alg:MP} to solve the problem formulated in \Stepref{step:formProblem} and to obtain a step $\begin{bmatrix} \Delta x^\Transp & \Delta \bar \theta^\Transp \end{bmatrix}^\Transp$ for the time ordered problem or a step $\begin{bmatrix} \Delta x^\Transp & \Delta \theta^\Transp \end{bmatrix}^\Transp$ for the sensor / segment ordered problem.}
\State{Update $x := x + \Delta x$, $\theta := \theta + \Delta \theta$ or $\bar \theta := \bar \theta + \Delta \bar \theta$.}
\State{Set $k := k+1$ and check for convergence and feasibility.}
\EndWhile
\end{algorithmic}
\end{algorithm}

\section{Results and discussion}
\label{sec:results}
\begin{figure}
\begin{center}
\includegraphics[width = 0.35\columnwidth]{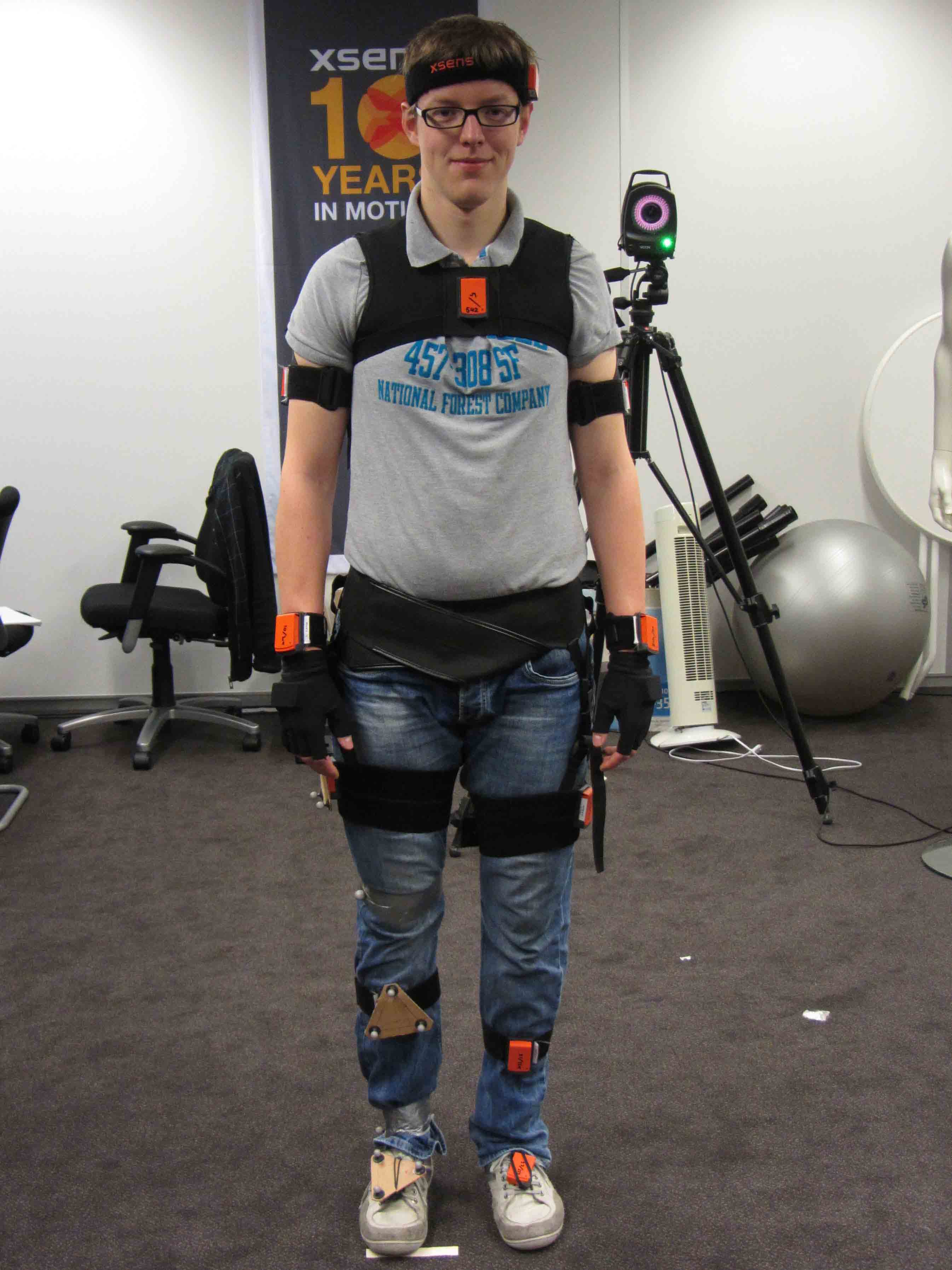}
\includegraphics[width = 0.35\columnwidth]{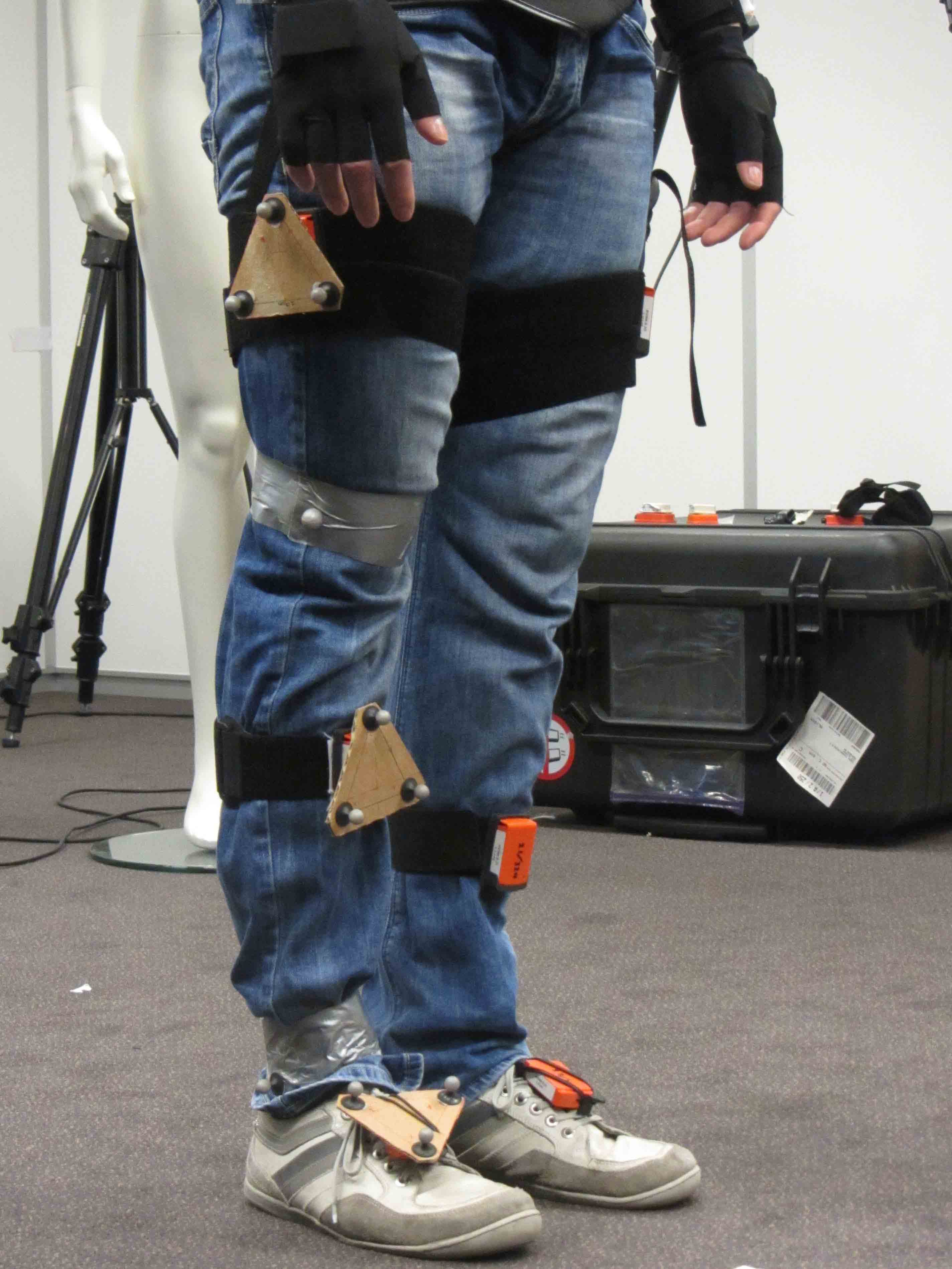}
\caption{\small Experimental setup where the human body is equipped with inertial sensors (orange boxes) on different body segments. High-accuracy reference measurements were obtained using an optical tracking system to validate the estimated joint angles. To this end, triangles with optical markers were placed on a number of sensors.\normalsize}  
\label{fig:experimentalSetup}                                 
\end{center}                                 
\end{figure}

We consider experimental data from a subject walking around for approximately 37 seconds wearing inertial sensors as shown in \Figureref{fig:experimentalSetup}. We focus on estimating the pose of the lower body using data from 7 sensors attached to the different body segments as described in \Sectionref{sec:model}. The estimated joint angles from this problem have previously been presented in~\cite{kokHS:2014}. In this work, we solve the same optimization problem but reorder the problem to efficiently make use of its structure. Hence, the estimates obtained using \Algorithmref{alg:resultingAlgorithm} are equal to the ones presented in~\cite{kokHS:2014}.

The optimization problem is solved at 10 Hz with $N_T = 373$, leading to a total number of $40284$ time-varying variables $x$ and $21$ constant variables $\theta$ and $6714$ constraints.\footnote{Note that the inertial sensors themselves are sampled at a much higher rate but strapdown integration \cite{savage:1998a,savage:1998b} is used to capture the high frequency signals, allowing for a lower update frequency of the optimization problem.} Notice that if the inherent sparsity of the problem would not be exploited, the computational complexity of solving the SQP for the smoothing problem  \eqref{eq:optimizationProblemTimeOrderedReformQP} or \eqref{eq:optimizationProblemSensorOrderedCompactQP} would grow cubically with the number of sensors and body segments $N_S$ and with the number of time steps $N_T$. The storage requirements for forming this problem would grow quadratically with $N_T$ and $N_S$.

To solve the problem in a more scalable manner, we have reordered the variables based on time and formed the problem as in~\eqref{eq:optimizationProblemTimeOrderedReformQPReorderRegroup}, which allows us to solve the problem using \Algorithmref{alg:resultingAlgorithm}. For each iteration $k$ in \Algorithmref{alg:resultingAlgorithm}, we then form $N_T - 1$ subproblems. Computing the messages in the upward pass requires each agent except the root agent to factorize a matrix of size $168$ since $|x_t| + |\bar \theta_t| = 129$ and $39$ constraints are involved in the subproblem. The root agent needs to instead factorize a matrix of size $315$ since $2 |x_t| + 2 |\bar \theta| = 258$ and $57$ constraints are involved in this subproblem instead. Using message passing to solve the problem, it is no longer required to form and store the large problem of size $46998$. Instead, it is only required to store one of these subproblems.

We have also solved the problem by reordering the variables based on sensors and segments. The computational benefits of this reordering are much less significant -- the problem is split up in 6 subproblems. However, the approach no longer requires collecting all data at a centralized unit, which can be communication intensive, and hence can potentially hamper our ability to have a seamless solution for the motion capture problem. Instead, it allows for decentralized computation of the solution, where the computational power on the sensors can be used to compute solutions to the subproblems.

\section{Conclusions and future work}
In this work, we have introduced a method to exploit the structure inherent in the inertial motion capture problem. The method allows for a scalable solution where small subproblems for each time step are formed and hence longer data sets can be processed. The approach is successfully applied to experimental data to estimate the pose of the lower body. It also opens up for the possibility of distributedly solving the problem by making use of the computational resources of each of the sensors. The structure that we exploit in this work is not unique to the motion capture problem. We believe that the message passing algorithm can be applied to a large number of other problems appearing in signal processing and estimation, e.g., in large-scale signal processing and estimation application. This is because these problems commonly enjoy desirable sparsity structures arising from physical and / or dynamic properties in the problem, as we saw for the inertial motion capture problem in this work.

\section*{Acknowledgment}
This work is supported by CADICS, a Linnaeus Center
and by the project Probabilistic modeling of dynamical systems
(Contract number: 621-2013-5524), both funded by the Swedish
Research Council (VR) and by the Swedish Department of Education within the ELLIIT project.

\bibliographystyle{unsrtnat}
\bibliography{literature}

\end{document}

%% file: coverArXiv.tex
\newcommand{\coverTitle}{A Scalable and Distributed Solution to the Inertial Motion Capture Problem}
\newcommand{\coverAuthors}{Manon~Kok, Sina~Khoshfetrat~Pakazad, Thomas~B.~Sch\"on, Anders~Hansson and Jeroen~D.~Hol}
\newcommand{\coverYear}{2016}

\begin{titlepage}
\begin{center}
%

\vspace*{2.5cm}
%
{\Huge \bfseries \coverTitle  \\[0.4cm]}

%
{\Large \coverAuthors \\[2cm]}

\renewcommand\labelitemi{\color{red}\large$\bullet$}
\begin{itemize}
\item {\Large \textbf{Please cite this version:}} \\[0.4cm]
\large
\coverAuthors. \coverTitle. In \textit{Proceedings of the
$\mathit{19}$th International Conference on Information Fusion},
pp. 1348--1355, Heidelberg, Germany, July 2016. 
\end{itemize}
\vfill

\begin{abstract}
In inertial motion capture, a multitude of body segments are equipped with inertial sensors, consisting of 3D accelerometers and 3D gyroscopes. Using an optimization-based approach to solve the motion capture problem allows for natural inclusion of biomechanical constraints and for modeling the connection of the body segments at the joint locations. The computational complexity of solving this problem grows both with the length of the data set and with the number of sensors and body segments considered. In this work, we present a scalable and distributed solution to this problem using tailored message passing, capable of exploiting the structure that is inherent in the problem. As a proof-of-concept we apply our algorithm to data from a lower body configuration.
\end{abstract}

\vfill
\end{center}
\end{titlepage}